\documentclass{aa}  

\usepackage{graphicx}
\usepackage{txfonts}
\begin{document}

   \title{Effects of XUV radiation on circumbinary planets}

   \author{J. Sanz-Forcada\inst{1}
          \and
          S. Desidera\inst{2}
          \and
          G. Micela\inst{3}
          }

   \institute{Departamento de Astrof\'{i}sica,
     Centro de Astrobiolog\'{i}a (CSIC-INTA), ESAC Campus, P.O. Box 78, 
     E-28691 Villanueva de la Ca\~nada, Madrid, Spain; \\
     \email{jsanz@cab.inta-csic.es}
     \and
     INAF -- Osservatorio Astronomico di Padova
     \email{silvano.desidera@oapd.inaf.it}
     \and
     INAF -- Osservatorio Astronomico di Palermo
     G. S. Vaiana, Piazza del Parlamento 1, Palermo, I-90134 Italy
     \email{giusi@astropa.inaf.it}
             }

   \date{Received ; accepted }

 
  \abstract
   {Several circumbinary planets have recently been discovered. The orbit
of a planet around a binary stellar system poses several dynamic
constraints. In addition to these constraints, the effects
that radiation from the host stars may have on the planet
atmospheres must be considered. We here evaluate these effects. 
Because of the
configuration of a close binary system, these stars have a high rotation
rate, even for old stars. The fast rotation of close, tidally
locked binaries causes a permanent state of high stellar activity
and copious XUV radiation. The accumulated effects are stronger than
for normal exoplanets around single stars, and cause a faster
evaporation of their atmospheres.}
   {We evaluate the effects that stellar radiation has on
       the evaporation of exoplanets around binary systems and on the
       survival of these planets.}
   {We considered the X-ray and EUV spectral ranges (XUV, 1--912 \AA) 
to account for the photons that are easily absorbed by a planet
atmosphere that is mainly composed of hydrogen. A more complex atmospheric
composition is expected to absorb this radiation more efficiently. We used direct X-ray observations to evaluate the energy in the X-rays range and coronal models to calculate the (nondetectable) EUV part of the spectrum.}
   {We considered in this problem different configurations of stellar
masses, and a resonance of 4:1 and 3:1. 
The simulations show that exoplanets
orbiting close binary systems in a close orbit will suffer 
strong photoevaporation that may cause a 
total loss of atmosphere in a short time. We also applied our models to
the best real example, \object{Kepler-47~b}, to 
estimate the current mass-loss rates in circumbinary planets and the
accumulated effects over the time.}
   {A binary system of two solar-like stars will
be highly efficient in evaporating the
atmosphere of the planet (less than 6 Gyr in
our case). These systems
will be difficult to find, even if they are
dynamically stable. Still, planets may orbit around binary systems of
low mass stars for wider orbits. Currently known
circumbinary planets are not substantially affected by thermal
photoevaporation 
processes, unless \object{Kepler-47~b} has an inflated atmosphere. The
distribution of the orbital periods of circumbinary planets is shifted to much
longer periods than the average of Kepler planets, which supports a
scenario of strong photoevaporation in close-in circumbinary planets.
}


   \maketitle
%

\section{Introduction}

\begin{figure}[t]
  \centering
  \vspace{0.5cm}
  \includegraphics[width=0.40\textwidth]{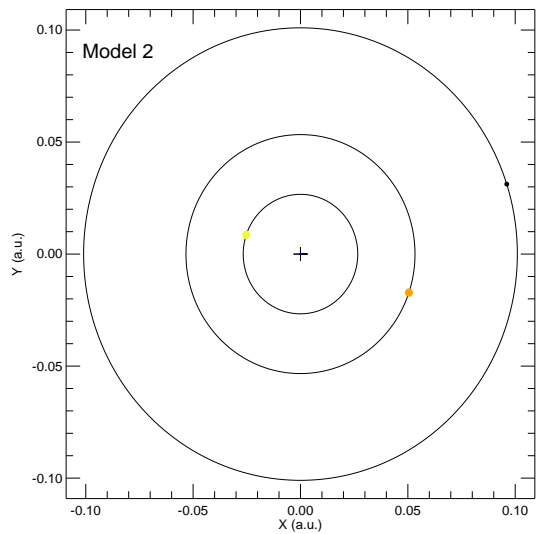}
  \caption{Orbit of model 2 (0.5 M$_\sun$ + 1.0 M$_\sun$). The planet size is not to scale ($M_{\rm p}$=1~M$_{\rm J}$).} 
  \label{fig:orbits}
\end{figure}

The recent discovery of several circumbinary planets
\citep[e.g.,][]{doy11,oro12} opens an
interesting debate on the formation and evolution mechanism of these
planets and the possible diversity of their properties compared with
planets orbiting single stars. A close binary
companion (separation $\la$ 100 a.u.) around a solar-type star was
shown to alter both the frequency and the characteristics of giant
planets \citep{bon07,des07}.
The orbit of a planet around a binary stellar system poses several dynamic
constraints. Several works have analyzed the long-term stability of such
a system \citep[][and references therein]{dvo89,hol99}. 
A recent study \citep{sch11} analyzed
the dynamical stability of circumbinary planets using the models
described in Fig.~\ref{fig:orbits}. There are, in general, three types of
orbits in which we can find a planet around two stars:
\begin{itemize}
\item S-Type: the planet orbits one of the stars in a wide binary system (e.g., \object{GJ 86 b}).
\item P-Type: the planet orbits the entire binary (e.g., \object{Kepler 35 b}).
\item L-Type: the planet is close to the Lagrangian points L4 or L5
(Trojan planets), as in some satellites in the solar system.
\end{itemize}

%
%

The analysis of the dynamical stability yields restrictions on
different variables of the problem. We focus here on
P-type systems. The binary system variables, such as semimajor axis,
stellar masses, or eccentricity, impose a minimum value of the
semimajor axis that allows the stability of the planet orbit in the
long term \citep{hol99}. In particular, planets with a 3:1 resonance
with the binary system ($P_{\rm orb}=3\, P_{\rm bin}$ or $a_{\rm p}=2.08\,
a_{\rm bin}$) 
are unlikely to be stable in the long term, even for circular orbits.
The eccentricity of the planet, combined 
with the stellar masses, imposes additional restrictions on the planet
semimajor axis: the closest planets can be found in 
systems with lower eccentricity and stars with similar
masses. An eccentricity higher than 0.3 hampers the survival
of a close-in circumbinary planet \citep{sch11}, although such planets
may exist in more distant orbits. 
A close central binary system is also expected to exert relevant perturbations
of the surrounding circumbinary disks, with strong implications for planet
formation. Planet formation is most probable far from the inner edge of
a circumbinary disk \citep{mar13}. In the inner regions 
the relative velocities between planetesimals are expected to prevent
accretion \citep{paa12,marz13}, although 
in-situ formation cannot be completely ruled out in special 
conditions \citep{mes13}. 
On the other hand, the preliminary statistical estimates based on the Kepler
data set indicate that circumbinary planets are moderately frequent
\citep[3\% considering short-period planets alone,][]{wel13}.

In addition to the dynamical constraints, the effects of
radiation from the host stars on the planet atmospheres need to be
considered. In this work we evaluate these effects. Planet gases 
will absorb energetic photons and eventually cause evaporation. In the
long term, this evaporation may result 
in the total stripping of the planet atmosphere, leaving behind only a
central rocky core with a small atmosphere. The mass-loss rate of the
planet mainly depends on the planet composition and the radiation
received. A high-density planet will resist evaporation longer,
as it will do an atmosphere with heavier species. We here
consider the simplest case, an atmosphere dominated by
hydrogen. Photons in the X-rays and EUV spectral ranges (XUV, 1--912
\AA) will be easily absorbed by such a planet. A more complex
atmospheric composition is expected to absorb this radiation more efficiently,
which does not necessarily cause a larger mass loss.  
We then assume that all the
absorbed energy causes atmospheric evaporation, although different
degrees of efficiency may be considered \citep[see][and references
  therein]{hub07,pen08,san11,owe12}. Additional effects such as nonthermal
losses due to 
stellar winds \citep{lan13} are not considered in this work. 

In the standard evolution of a single star the coronal radiation is
typically highest ($L_{\rm X}\sim 10^{-3} L_{\rm bol}$) at young ages
(10--100 Myr), when the high rotation yields a high level of activity.
Then it gradually fades as the star spins down
during its evolution \citep[e.g.,][]{fav03}. In close
binary systems with periods
shorter than a few days, tidal interaction forces 
the stars to maintain a high rotation
rate, even for old stars. The fast rotation of close, tidally
locked binaries, causes a permanent state of high stellar activity
\citep[e.g., RS CVn stars,][]{wal78}, 
and a copious XUV radiation. 
Chromospheric lines are typically observed in emission in these
cases, including \ion{Ca}{ii}\,H\&K, the Ca infrared triplet (IRT),
and H$\alpha$ \citep[e.g.,][]{str88,mon00}. The stellar corona is more
sensitive to activity, with copious emission in the 
X-rays and extreme ultraviolet (EUV) bands in close binary systems
\citep{pal81,dem93,dup93,san03}.
The accumulated effects are stronger than
for exoplanets orbiting single stars at the same distance of the parent
star, and cause a faster evaporation of their atmospheres. Evidence of
atmospheric evaporation has been reported in planets around single
stars \citep[][and references therein]{san10}. Stellar XUV radiation
decays quickly over time in single stars, therefore it is expected that
effects are stronger in planets around close binaries.

We carried out numerical simulations with different configurations of stellar
masses to test these effects (Sect. 2), with early results presented
in \citet{san12}. In Sect. 3 we apply our
simulations to the currently known circumbinary systems to estimate
the current mass-loss rates in 
circumbinary planets and the 
accumulated effects over time. In Sect. 4 we discuss the results
and consider the minimum
distance at which a circumbinary planet may survive; we conclude 
in Sect. 5.

%
\begin{figure*}[t]
  \centering
  \vspace{0.5cm}
  \includegraphics[width=\textwidth]{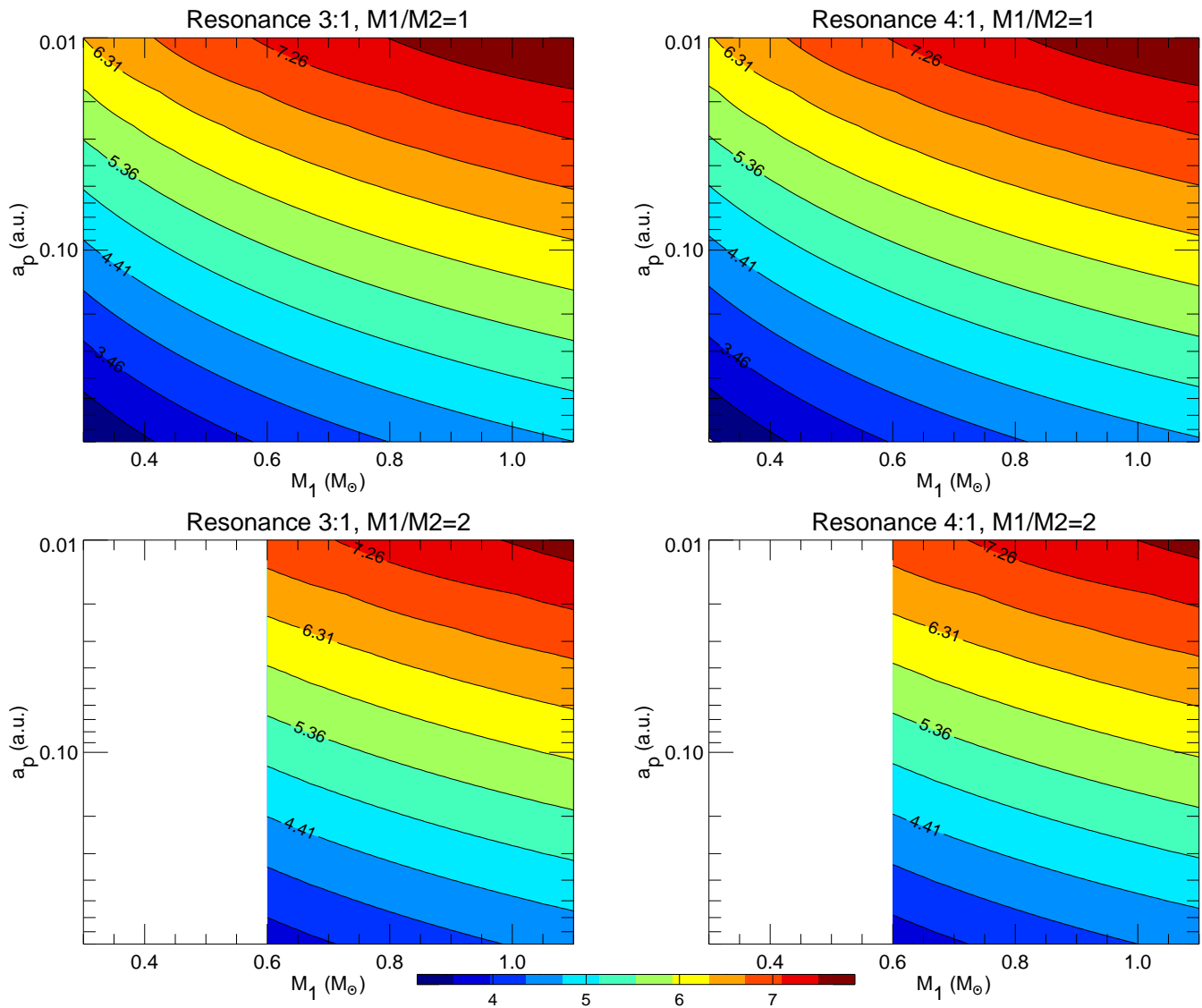}
  \caption{Stellar XUV flux ($\log F_{\rm XUV}$, in
    erg\,s$^{-1}$\,cm$^{-2}$) 
    received at the planet orbiting a binary system under activity
    saturation regime (color coded). Resonance 3:1 and 4:1 are considered. Upper
    panel shows the case of $M_1=M_2$, while lower panel uses 
    $M_1=2\,M_2$. The planet HD 189733 b, orbiting a single star,
    receives $\log F_{\rm XUV}$=4.2 erg\,s$^{-1}$\,cm$^{-2}$ \citep{san11}.
}
  \label{fig:contornos}
\end{figure*}
%

\section{Numerical simulations}
The mass-loss rate due to atmospheric evaporation depends on
the high-energy radiation received. Before we select a model for a mass-loss
rate we need to evaluate $F_{\rm XUV}$, the X-ray and EUV
flux received at the planet position.
For this purpose we considered the general case of a planet orbiting a binary
stellar system, without eccentricity in either orbit, and in resonant
orbits (the ratio $P_{\rm orb}/P_{\rm bin}$ is an integer). The flux
received at the planet in that case is (see Appendix~\ref{app:formulas})

\begin{equation}\label{eq:flux}
F=\frac{1}{4\pi}\, \Bigg(\frac{L_{\rm XUV,1}}{a_p^2-a_1^2}+\frac{L_{\rm
    XUV,2}}{a_p^2-a_2^2}\Bigg)\, ,
\end{equation}
\noindent
where $L_{\rm XUV,1,2}$ is the XUV luminosity of each star, $a_p$ is
the semimajor axis of the planetary orbit, and $a_1$ and $a_2$ are
related to the binary semimajor axis and the stellar masses following
\begin{displaymath}
a_1 = \frac{2a}{1+\frac{M_1}{M_2}}\, ,\quad a_2 =
\frac{2a}{1+\frac{M_2}{M_1}}\, .
\end{displaymath}

We assumed that rotational and orbital periods of the two stars are
synchronized, 
forcing a high rotation rate, as is usual in tidally locked stellar
binaries. Tidal locking is favored by small stellar separation, and
it yields a faster spin until it couples the orbital 
period, on a time scale that is shorter for bodies of similar masses
\citep[][and references therein]{zah77,zah89,sch00}. The
synchronization time has a dependence of the type 
$\tau_{\rm sync}\sim (\frac{M_1}{M_2})^2\,(\frac{a}{R})^6$.
Most
main-sequence spectroscopic binaries tend to be synchronized for
periods shorter than eight days.
The main effect of a faster rotation is a strong
dynamo that causes a high activity level. If the rotation rate
continues to be
high, the activity level will remain stable. 
We here assumed that the rotation is fast enough to keep the X-ray
emission at the saturation regime, when the X-ray flux is
highest\footnote{Saturation in the X-ray flux takes place for rotation
periods shorter than about 2--3~d for G stars, 3.3~d for K stars, and more than
10~d for M stars \citep{piz03}.}. 
The XUV luminosity of a star in the saturation regime can be
related to stellar mass ($0.2 \la M \la 1.1$ M$_\sun$) with
this approximation (see Appendix~\ref{app:formulas}):
\begin{equation}\label{eq:luminosity}
\frac{L_{\rm
    XUV}}{L_\sun}=7.57\times10^{-4}\,\Bigg(\frac{M}{M_\sun}\Bigg)^{3.8}+130.69\,\Bigg(\frac{M}{\rm{M}_\sun}\Bigg)^{3.268}\, .
\end{equation}
Figure~\ref{fig:contornos} displays the flux received by a planet
orbiting a binary system with mass ratios $M_1/M_2$ 1 and 2,
and resonance 3:1 and 4:1 (see below). As a reference, the exoplanet
HD~189733 and Earth receive a $\log F_{\rm
  XUV}$(erg\,s$^{-1}$\,cm$^{-2}$)$\sim$4.2 and 0.5 respectively
\citep{san11}. 
In currently known circumbinary systems
(e.g., Kepler-47 AB, with $P_{\rm orb}=7.45$~d, Table~\ref{tab:zoo}) the
activity is most likely 
lower than this \citep[see][]{piz03}, therefore assuming of a
saturation regime places an upper limit on the photoerosion of the planet. 

We simulated the radiation effects for different
configurations of the stellar binary system.
We studied the
evaporation of a Jupiter-like planet with a density of $\rho$=1~g\,cm$^{-3}$
($\rho_{\rm J}$=1.24~g\,cm$^{-3}$). Erosion and time evolution of the XUV 
luminosity from \citet[][and references therein]{san11} are
used in the calculations. The mass-loss rate due to thermal
photoevaporation in a hydrogen planet atmosphere is calculated
as 

%
\begin{table*}
\caption[]{Planets orbiting a binary system}\label{tab:zoo}  
\begin{center}
\begin{small}
\vspace{-6mm}
\begin{tabular}{lcccccccccccc}
\hline \hline
Planet name & $M_{\rm A/B}$ & $a_{\rm bin}$ & $a_{\rm p}$ & $M_{\rm
  p}$\tablefootmark{a} & $e_{\rm p}$ & $P_{\rm orb}$ & $P_{\rm bin}$ &
$e_{\rm bin}$ & $a_{\rm crit}$ & $a_{\rm p}$/$a_{\rm crit}$ &
$d_{\rm periastron}/a_{\rm crit}$ \\
  &  M$_\sun$  &  a.u. & a.u. & M$_{\rm J}$ &  &  & d &  d &  a.u. & & \\
\hline
\object{Kepler 16 b} & 0.69/0.20 & 0.224 & 0.7048 & 0.333 & 0.007 & 228.776 & 41.08 & 0.15944  & 0.646 & 1.09 & 1.08 \\
\object{Kepler 34 b} & 1.05/1.02 & 0.229 & 1.0896 & 0.220 & 0.182 & 288.822 & 27.80 & 0.52087 & 0.836 & 1.30 & 1.06 \\
\object{Kepler 35 b} & 0.89/0.81 & 0.176 & 0.6035 & 0.127 & 0.042 & 131.458 & 20.73 & 0.1421 & 0.497 & 1.21 & 1.16 \\
\object{Kepler 38 b} & 0.95/0.25 & 0.1469& 0.4644 & (0.076)& <0.032 &105.595 & 18.80 &0.1032  &0.389  & 1.19 & N/A \\
\object{Kepler 47 b} & 1.04/0.36 & 0.0836& 0.2956 & (0.024)& <0.035  & 49.514 & 7.45 &0.0234  &0.202  & 1.46 & N/A \\
\object{Kepler 47 c} &           &       & 0.989  & (0.090)& <0.411  & 303.148 &  &  &  & 4.90 & N/A \\
\object{PH1 b}          & 1.53/0.41 & 0.1744 & 0.634 & 0.53 & 0.054 & 138.506 & 20.00 & 0.2165 & 0.533 & 1.19 & 1.13 \\
\object{Kepler 413 b} & 0.82/0.54 & 0.1015 & 0.355 & 0.211 & 0.118 & 66.262 & 10.12 & 0.037 & 0.260 & 1.37 & 1.20 \\
\hline
\end{tabular}
\end{small}
\end{center}
\vspace{-6mm}
\tablebib{\citet{doy11}, \citet{wel12},
    \citet{oro12b}, \citet{oro12}, \citet{sch13}, \citet{kos14}}
\tablefoot{
\tablefoottext{a}{Masses in parenthesis were calculated using the
  planet radius and the density of Neptune (1.61\,g\,cm$^{-3}$).}
}
\end{table*}

\begin{equation}\label{eq:massloss}
\dot M=\frac{3 F_{\rm XUV}}{4\, {\rm G} \rho}\, ,
\end{equation}
\noindent
where G is the gravity constant, and $\rho$ is the planet
density\footnote{An evolution with  constant density was assumed. This
  assumption does not substantially affect the conclusions
  \citep{san11}.}. 
This formula represents a quite simple approach to the
thermal losses and probably does not truly represent the actual evaporation
rate (detailed models predict an efficiency of
  $\la$10\%). In addition, radiation levels exceeding  
$10^5$~erg\,s$^{-1}$\,cm$^{-2}$ might slow down the increase rhythm of
the evaporation rates \citep{mur09}. But it serves the purpose of a
qualitative modeling of the problem. Finally,
nonthermal losses due to
stellar winds \citep{lan13} can be of the same order as the thermal
losses considered here. The effect of these winds might
roughly double the mass-loss rate in these exoplanets. We did not
include these effects here.

To simulate the mass loss of the planets for this simple approach
we considered three different models depending on the stellar masses, 
with the same configurations as in \citet{sch11}:
0.3 M$_\sun$ + 0.3 M$_\sun$ (model 1), 0.5 M$_\sun$ + 1.0 M$_\sun$
(model 2), and 1 M$_\sun$ + 1 M$_\sun$ (model 3). In all cases the
semimajor axes were set to $a$=0.04 and 
$a_{\rm p}$=0.101. This configuration results in a 4:1
resonance, to avoid conflict with the long-term instability of the
system that might be present with a resonance 3:1.
An eccentricity of $e$=0.0 was
used in the simulations, consistently with the hypothesis of
tidally locked binaries. The currently known planet
with the closest orbit is \object{Kepler-42~c}
\citep[$a$=0.006,][]{mui12}, and the 
circumbinary planet with shortest distance is Kepler-47~(AB)~b
($a_{\rm p}$=0.2956, Table~\ref{tab:zoo}). 

\begin{figure}[t]
  \centering
  \vspace{0.5cm}
  \includegraphics[width=0.5\textwidth]{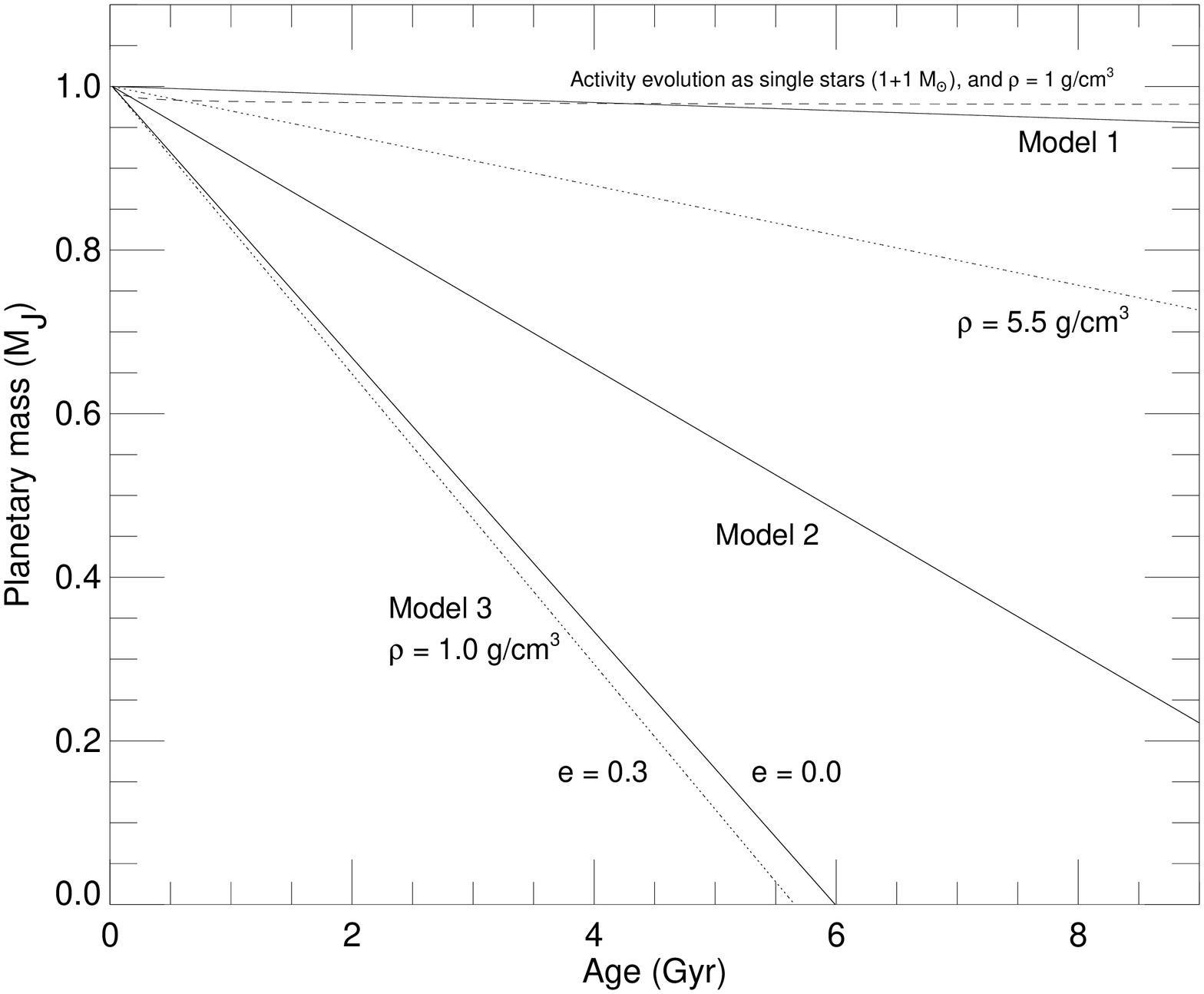}
  \caption{Mass evolution for the different scenarios for an initial
mass of 1 M$_{\rm J}$, assuming a constantly high activity
level. Solid lines indicate the models with an 
eccentricity 0.0 and a density of 1 g cm$^{-3}$ (model 1: 0.3 M$_\sun$ +
0.3 M$_\sun$. model 2: 0.5 M$_\sun$ + 1.0 M$_\sun$. model 3: 1
M$_\sun$ + 1 M$_\sun$). The effects on an Earth-like planet would be devastating
even if it orbits a two M-star system. Dotted lines are used for variations of
density and eccentricity in model 3. A higher density (Earth
density is 5.5 g cm$^{-3}$) protects the planet 
against photoevaporation. A dashed line indicates 
the evolution with the usual activity decay over time of single stars.
}
  \label{fig:comparison}
\end{figure}

We calculated the effects of the stellar XUV flux evolution for a
whole planetary period (15.12~d in 
model 1, 9.57~d in model 2, and 8.29~d in model 3)
assuming constant XUV stellar luminosity during the period.
Then we convolved the effects with the age of the
system, starting at an age of 20~Myr, when most circumstellar disks are
dissipated, and ending at 13.7~Gyr (approximate age of
the Universe). This value was reduced to 9~Gyr when there is at least
one solar-mass star in the system, corresponding to its approximate
evolution lifetime in the main sequence.
As an example, the range of XUV flux received at the planetary orbit
in model 2 (resonance 4:1) is $\log F_{\rm XUV}$~[erg s$^{-1}$
  cm$^{-2}$]=5.6--5.9, 
with secondary maxima at  $\log F_{\rm XUV}$=5.7 because of the
different flux from the two stars.
Fig.~\ref{fig:comparison} displays the simulation results for the
three models. The highest photoevaporation is seen in the model 
3 (1 M$_\sun$ + 1 M$_\sun$). The absolute XUV flux from the solar-like
stars is much higher than that of the M stars because of their size,
resulting in a fast ($\sim$6 Gyr) evaporation of the jovian
planet. The planet around two M stars would survive in the long term,
although the mass 
lost is considerable and would severely affect a planet of smaller
mass. 

In Fig.~\ref{fig:comparison} we also show the effects in model 3 for
a higher eccentricity ($e$=0.3) and a different planet density. 
We include a comparison with the normal evolution of
coronal emission in single late-type stars 
\citep[$L_{\rm X}\sim  t^{-1.55}$, $L_{\rm EUV}\sim t^{-1.24}$ 
after the saturation period,][]{san11}. An eccentricity of 0.3 would result in
only a slightly stronger photoerosion effect. A higher planet density
substantially protects it from photoerosion, as expected from
Eq.~\ref{eq:massloss}. The same stellar configuration, but with a
normal evolution of the stellar activity would result in very weak
effects for a jovian planet. But currently known jovian exoplanets include
cases with a density as low as $\sim$0.1~g\,cm$^{-3}$; in this case,
the planet would loose the whole atmosphere in less than 4 Gyr even in
model 1.

\section{Application to current circumbinary planets}
We tested the scenario of planetary mass loss due to radiation on real
cases. Table~\ref{tab:zoo} lists the sample of currently known
P-type circumbinary planets. We excluded from the list the massive
planets or brown dwarfs in a
wide orbit ($a_{\rm p}>$80~a.u.) that were discovered by direct imaging
\citep[Ros~458~b, SR~12~c, 2MASS0103, FW Tau~b and ROXs
  42B~b;][]{bur10,kuz11,del13,kra14}, because the large separation 
from the central stars makes the effects discussed here absolutely
negligible. We also excluded binary 
systems with cataclysmic variables or white dwarfs
\citep[PSR B1620-26, HU Aqr, HW Vir, NN Ser, DP Leo, UZ For, RR Cae, and NY Vir;][]{sig03,hin12,beu10,beu11,beu12,pot11,qia12,qia12b},
because the common-envelope evolution 
of the central system would cause a different astrophysical situation.  
\object{Kepler-47~b} is the circumbinary planet with the closest
distance to the binary system (Table~\ref{tab:zoo}) and receives the
highest amount of XUV radiation.  
\object{Kepler-35~b} orbits a pair
with higher emitted flux (because of larger stellar size) than
\object{Kepler-47}, but this is not enough to compensate for its larger
semimajor axis.

\begin{figure}[t]
  \centering
  \vspace{0.5cm}
  \includegraphics[width=0.5\textwidth]{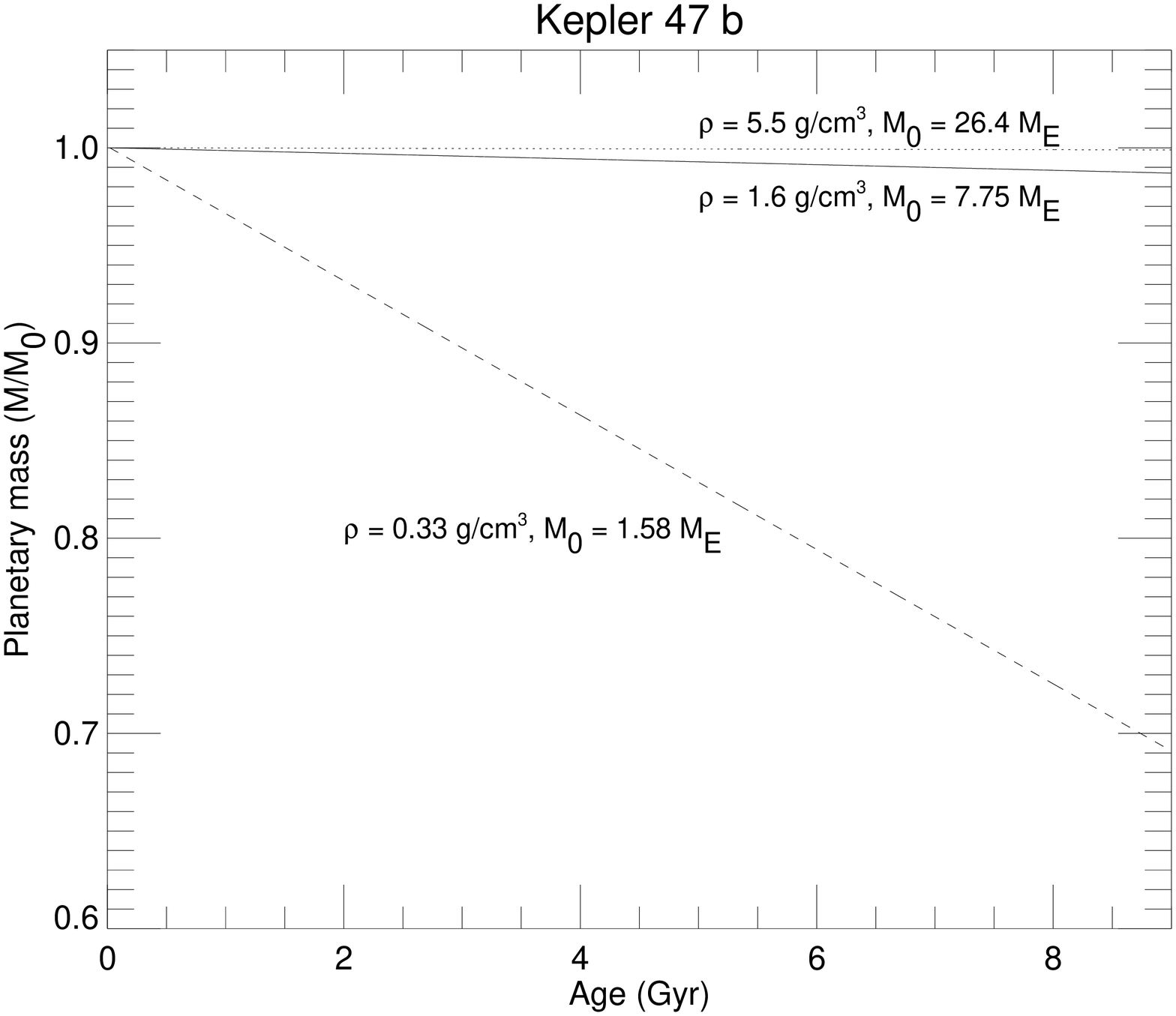}
  \caption{Mass evolution of \object{Kepler-47~b} with age with respect to
    initial mass for three different densities (mass expressed in
    Earth units). The solid line marks a Neptune-like
    density. Dashed and dotted lines indicate the evolution in extreme
    cases of density. This is the most favorable
    system discovered to date for the scenario of planet
    photoevaporation.
}
  \label{fig:kepler47}
\end{figure}

For the test we assumed that the orbit of the binary system Kepler 47
AB is close enough to force a tidal
synchronization between orbital and rotational
periods, thus yielding an XUV emission close to its maximum (an
emission of $\log L_{\rm XUV}=30.98$ is assumed in
\object{Kepler-47~b}). This will give us an upper limit of the mass-loss
rate. The results (Fig.~\ref{fig:kepler47})  
show that the impact of XUV radiation on the mass loss depends much on
the (unknown) mass of the planet.
Assuming the density of Neptune\footnote{The density of
  Neptune-size exoplanets is very variable.}
($\rho_{\rm N}$=1.61~g\,cm$^{-3}$), which is
appropriate for the mass of \object{Kepler-47~b}, 1.3\% of the original
mass is lost after 9 Gyr (average $\dot M=2.12\times10^{9}$g\,s$^{-1}$). 
The effects are stronger for
a lower mass planet, although a higher density (typical
of Earth-like planets) would also protect it. Therefore we added two
different scenarios based on the values observed in currently known
planets with similar radii. Most of them have a mass between 1.6 and
27 M$_\oplus$ ($\rho=0.33-5.5$~g\,cm$^{-3}$), which discards too high
values that would correspond to planets with little or no
atmosphere. In the lowest density case the planet losses
a remarkable $\sim$31\% of its mass.

%
%

Dynamical stability in the long term due to the configuration of the
binary system can be tested using the formula of \citet{hol99}. We
calculated the critical semimajor axis that allows the dynamical
stability of the planet in the long term. We then compared it with the
observed planet semimajor axis and its periastron for
the cases in which the 
eccentricity has been measured. All the cases in Table~\ref{tab:zoo}
have both quantities quite similar to the observed semimajor axis of the
planet, but still lower than it.

\section{Discussion}
Currently known circumbinary systems
are not subject of substantial photoerosion by XUV radiation from
central stars, unless they have a rather inflated atmosphere.
To test whether we can expect
to find circumbinary planets exposed to higher radiation, we considered
the shortest distance at which a planet
can survive the effects of radiation, depending on the stellar and
planetary masses, and the density of the planet. 
The evaporation rates from Eq.~\ref{eq:massloss} were used as a
first approximation to the problem.

We considered the three models of stellar masses 
(1+1~M$_\sun$, 0.5+1~M$_{\sun}$ and 0.3+0.3~M$_{\sun}$) and two
types of planets: a jovian planet (1~M$_{\rm J}$) and a
neptunian planet (0.054\,M$_{\rm J}$ or
17.15\,M$_{\oplus}$). Experimental data show that 
jovian-mass planets have densities below 1.3~g\,cm$^{-3}$, and
close-in planets tend to be inflated, possibly because of
the XUV absorbed radiation \citep[e.g.,][but other explanations have been
  proposed]{san11}, therefore we used 
densities of 1.3, 0.8 and 0.3~g\,cm$^{-3}$. For the neptunian planet
we used 2, 1, and 0.5~g\,cm$^{-3}$ (Neptune has
$\rho$=1.6~g\,cm$^{-3}$). The results are displayed in
Table~\ref{tab:survival}. We include the results for the scenarios
with resonance 4:1 and 3:1.

The results show that planets with a short orbital period may suffer
the effects of radiation to a great extent, which causes its total
evaporation at a distance of 0.168 a.u. in 5 Gyr for an inflated Jupiter-like
planet (in resonance 4:1), depending on stellar mass. Neptunian
planets have even lower possibilities of survival 
(below $a_{\rm p}$=0.562 if $\rho$=0.5\,g\,cm$^{-3}$). There are two
limitations that need to be considered with this values: first, a
higher value of $a_{\rm p}$ also implies a larger semimajor axis in the
binary system ($a$=0.223 a.u. for the mentioned neptunian example),
which means that rotation and orbital motions are not as highly
coupled in the stars; 
the second is the long-term stability of the system, depending on the
critical value of the binary star's semimajor axis, eccentricity, and
masses. Equation 3 of \citet{hol99} reveals that P-type
planets in an orbit with resonance 3:1 ($a_{\rm p}$=2.08 $a_{\rm
  bin}$) are very unlikely to last in
the long term, while those with resonance 4:1 ($a_{\rm p}$=2.52
$a_{\rm bin}$) will last only if the eccentricity is close to
0. \citet{sch11} concluded that some planets with resonance 3:1 may
last long enough to be detected.

%

From the values in Table~\ref{tab:survival} we can conclude that 
circumbinary planets close to the dynamical stability limits around
short-period tidally locked binaries can exist, and they will be
subject to intense evaporation effects due to photoerosion.
The most massive
planets might survive longer. The shortest distances required for the
circumbinary planets are at least twice as long as those of the current
planet with the closest orbit (\object{Kepler-42 c}, $a_{\rm bin}$=0.006 a.u.),
depending also on the spectral type of the parent stars. 

\begin{figure}[t]
  \centering
  \vspace{0.5cm}
  \includegraphics[width=0.45\textwidth]{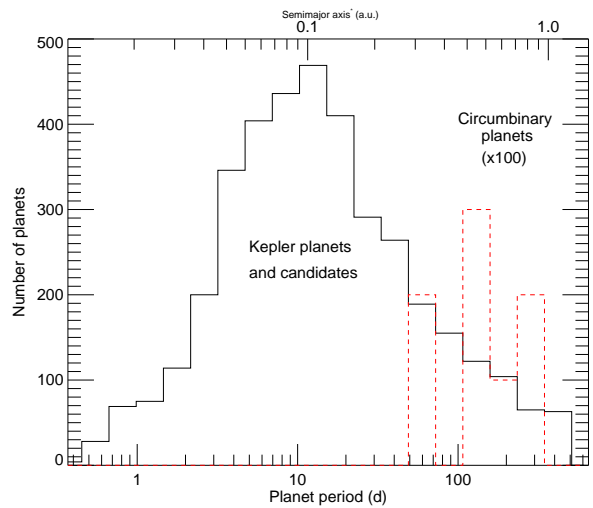}
  \caption{Distribution of Kepler planet candidates with orbital
    period (solid line). The dashed line indicates the distribution of
    the population (multiplied $\times100$) of known Kepler
    circumbinary planets. The upper axis marks the corresponding
      value of the semimajor axis calculated for a host star with 1~M$\sun$
    and this range of periods.
}
  \label{fig:periods}
\end{figure}

The effects of photoerosion imply that circumbinary planets with a
too small orbit will be difficult to find and will likely be rocky
planets, as an effect of the atmosphere stripping by the XUV
radiation. We can test the validity 
of this hypothesis by comparing the distribution of orbital periods of
planets discovered by Kepler with that of the circumbinary planets in
Table~\ref{tab:zoo}, all of which were discovered with Kepler
(Fig.~\ref{fig:periods}). 
It is remarkable to see that while most
exoplanets have a short orbital period, the Kepler
circumbinary planets have longer orbital periods in all cases. 
A difference in orbital period between circumbinary planets and
planets orbiting single stars is obviously expected considering the
stability limits that exclude the inner regions around a binary
system. The currently known circumbinary planets are
mostly found close (within 20\%) to the dynamical stability limits
(Table~\ref{tab:survival}). This is consistent with some theoretical
predictions of planet migration in circumbinary disks \citep{pie13}.
Furthermore, the detection of Kepler planets around single stars 
is biased toward those with a
short period \citep[$\la 200$~d,][]{pet13}. But the different
distribution 
of circumbinary planets in Fig.~\ref{fig:periods} is difficult to
explain with these effects alone, even considering the
small size of the sample in Table~\ref{tab:zoo}. In fact,
Kepler observed a large sample of short-period eclipsing binaries
\citep[][their Fig. 8]{sla11} and about half of the detached and
semi-detached binary systems have orbital periods shorter than 5~d. On
the other hand, all the eclipsing binary systems with circumbinary
planets discovered so far have an orbital period longer than 7~d.
This difference was also noted in \citet{wel13},
who did not ascribe this to selection effects, because close planets
should be easier to discover around closer binaries because of geometrical 
effects and more transits. On the other hand,
\citet{wel13} did not mention the effects of the
higher levels of magnetic activity of the closer binaries on planet 
detectability.
A more detailed study is required to asses whether selection effects
play a role in the currently observed 
lack of circumbinary planets around the shortest-period eclipsing
binaries.

%
\begin{table}
\caption[]{Smallest planet semimajor axis for its survival after 5 Gyr\tablefoottext{a}}\label{tab:survival}  
\tabcolsep 3 pt
\begin{center}
\begin{small}
\vspace{-6mm}
\begin{tabular}{l|ccc|ccc}
\hline \hline
 & \multicolumn{3}{c}{Jovian planet} & \multicolumn{3}{c}{Neptunian planet} \\ 
Stellar masses & $\rho=1.3$ & 0.8 & 0.3 & $\rho=2$ & 1 & 0.5 \\
\hline
 & \multicolumn{6}{c}{Resonance 4:1}\\
 1 M$_\sun$ + 1 M$_\sun$     & 0.081 & 0.103 & 0.168 & 0.281 & 0.397 & 0.562 \\
 0.5 M$_\sun$ + 1 M$_\sun$   & 0.058 & 0.074 & 0.121 & 0.202 & 0.286 & 0.404 \\
 0.3 M$_\sun$ + 0.3 M$_\sun$ & 0.014 & 0.018 & 0.029 & 0.048 & 0.068 & 0.096 \\
& \multicolumn{6}{c}{Resonance 3:1}\\
 1 M$_\sun$ + 1 M$_\sun$     & 0.085 & 0.108 & 0.176 & 0.294 & 0.416 & 0.588 \\
 0.5 M$_\sun$ + 1 M$_\sun$   & 0.060 & 0.076 & 0.125 & 0.208 & 0.294 & 0.416 \\
 0.3 M$_\sun$ + 0.3 M$_\sun$ & 0.014 & 0.018 & 0.030 & 0.050 & 0.071 & 0.101 \\
\hline
\end{tabular}
\end{small}
\end{center}
\vspace{-6mm}
\tablefoot{\tablefoottext{a}{Planet densities in g\,cm$^{-3}$}, planet
  semimajor axis in a.u.}
\end{table}

If it is real, a possible explanation which we are
  proposing here is that circumbinary
planets form or migrated preferentially close to the dynamical
stability limits \citep{mar13}, but they are strongly affected by
evaporation effects and then reduced to compact remnants, making their
detection challenging. 
An alternative hypothesis, also mentioned in \citet{wel13}, is that
the closest binary systems 
($P<3-5$~d) do not form circumbinary planets at all, possibly because
they very often have tertiary stellar companions \citep{tok06} that
may have played a role in the tightening of the orbit of the inner
binary \citep[e.g., through Kozai interactions, see ][]{maz79}. This
evolution is expected to be disruptive for circumbinary planets.
Future observations of circumbinary planets around close binaries
and a comprehensive statistical analysis on the frequency and properties
of circumbinary planets at different binary orbital periods will provide 
more information on the occurrence of evaporation on
their circumbinary planets, and by extension, to exoplanets with
close-in orbits to single stars. An observational problem is
measuring of the planetary mass because it is difficult to
determine of the radial velocity in circumbinary planets.

\section{Conclusions}

The simulations showed that exoplanets
orbiting close binary systems will suffer 
strong photoevaporation that may cause 
total loss of atmosphere in a short 
time, depending on the initial planet mass, because of the high levels of
magnetic activity that last the whole stellar lifetime.

A binary system of two solar-like stars will
be highly efficient in evaporating the
atmosphere of the planet. This indicates that these systems 
will be difficult to find, even if they are
dynamically stable.

Currently known circumbinary planets are not
affected by thermal photoevaporation processes, except in a scenario
of an inflated atmosphere in \object{Kepler-47~b}. The distribution of the
orbital periods of Kepler 
circumbinary planets shows much longer periods than the average of
Kepler planets, supporting the hypothesis of strong photoevaporation effects 
in close-in circumbinary planets. 

The thermal evaporation caused by radiation we evaluated here is only part of 
the problem. Nonthermal processes, such as ion-picking by stellar
winds, probably
increase the mass-loss rate of these planets and the effects in the 
long term on the planetary mass.

%

\begin{acknowledgements}
JSF acknowledges support from the
  Spanish MINECO through grant AYA2011-30147-C03-03. SD and GM
  acknowledge support from PRIN INAF 2010 
"Planetary systems at young ages and the interaction with their active 
host stars"
\end{acknowledgements}


\Online
\begin{appendix}

\section{Flux received by a planet around a binary star}\label{app:formulas}
To calculate the mass-loss rate due to evaporation (thermal losses) we
proceeded in two steps. First we derived the stellar flux
received by the planet in a given wavelength range (XUV in our case);
second, we applied a model that calculates how that flux
results in atmospheric evaporation, and subsequently in planet mass-loss. In
this appendix we include the necessary formulae for calculating the flux
received by a planet orbiting a close binary system composed of two
late-type stars. 

The flux received by a planet, averaged over a whole orbital period,
can be calculated from
\begin{equation}\label{eq:flux1}
F=\frac{1}{4\pi P_{\rm orb}} \int_{t=0}^{t=P_{\rm orb}}
\Bigg(\frac{L_1}{r_1^2} + \frac{L_2}{r_2^2}\Bigg)\, dt\, ,
\end{equation}
\noindent
where $L_{1,2}$ are the stellar luminosity in the considered wavelength
range, and $r_{1,2}$ are the distances between the planet an each of the
stars. Luminosities are considered constant during a period. $r_{1,2}$
are calculated by combining the movement of the stars in the system and
the planet itself, resulting in
\begin{equation}\label{eq:flux2}
r_{1,2} = a_{1,2}^2+a_p^2-2a_{1,2}a_p\;\cos(\phi_2-\phi_1)\, ,
\end{equation}
where $\phi_{1,2}= 2\pi \frac{t}{P_{\rm bin, orb}}$ and
$a_p$ is the semimajor axis of the planet's orbit.
$a_1$ and $a_2$ are
related to the binary semimajor axis and the stellar masses following
\begin{displaymath}
a_1 = \frac{2a}{1+\frac{M_1}{M_2}}\, ,\quad a_2 = \frac{2a}{1+\frac{M_2}{M_1}}\, .
\end{displaymath}
We can solve the integrals by using $x=\frac{2 \pi \, \big(\frac{P_{\rm
      orb}}{P_{\rm bin}}-1\big) t}{P_{\rm orb}}$ and assuming a resonant
orbit ($P_{\rm orb}/P_{\rm bin}$ is an integer):
\begin{equation}\label{eq:flux3}
\int_{t=0}^{t=P_{\rm orb}}\frac{dt}{r_1^2} \, =\, \frac{P_{\rm
    orb}}{2\pi a_p^2} \frac{2\pi}{1-(a_1/a_p)^2}\, .
\end{equation}

The resulting flux received at the planet's position is then

\begin{equation}\label{eq:flux4}
F=\frac{1}{4\pi}\, \Bigg(\frac{L_{\rm 1}}{a_p^2-a_1^2}+\frac{L_{\rm
    2}}{a_p^2-a_2^2}\Bigg)\, ,
\end{equation}
\noindent
where $L_{1,2}$ is the luminosity of each star in a given spectral
range.

We are interested in the XUV spectral range. The luminosity of a
late-type star under the saturation 
regime can be approximated by  \citep{san11}
\begin{equation}\label{eq:lum1}
L_{\rm XUV}=6.3\times10^{-4} L_{\rm bol}\, + \,111.54 L_{\rm bol}^{0.86}\, ,
\end{equation}
\noindent
which combined with the classical luminosity-mass relation \citep{cox00}
\begin{equation}\label{eq:lum2}
\log \frac{L_{\rm bol}}{L_\sun}=3.9 \log\,\Bigg(\frac{M}{M_\sun}\Bigg)+0.08
\end{equation}
\noindent
yields
\begin{equation}\label{eq:lum3}
\frac{L_{\rm XUV}}{L_\sun}=7.57\times10^{-4}\,\Bigg(\frac{M}{M_\sun}\Bigg)^{3.8}+130.69\,\Bigg(\frac{M}{\rm{M}_\sun}\Bigg)^{3.268}\, .
\end{equation}

This equation is valid in the range $0.2 \la M \la 1.1 M_\sun$.

\end{appendix}

\end{document}